\documentclass[twocolumn,showpacs,preprintnumbers,amsmath,amssymb]{revtex4-1}
\usepackage{graphicx}
\usepackage{dcolumn}
\usepackage{bm}
\usepackage{color}
\usepackage{enumitem,kantlipsum}
\usepackage{amsmath}
\usepackage{bm}
\usepackage[normalem]{ulem}
\usepackage{color}

\begin{document} 
\tighten
\newcommand {\be}{\begin{equation}}

\newcommand {\ee}{\end{equation}}

\newcommand {\bea}{\begin{eqnarray}}

\newcommand{\cred}{\color{red}}
\newcommand{\cmag}{\color{magenta}}
\newcommand{\cblue}{\color{blue}}

\newcommand{\cO}{{\cal O}}
\newcommand {\eea}{\end{eqnarray}}

\newcommand {\tq}{\theta_{{\bf q}}}

\newcommand {\bq}{{\bf q}}
\newcommand {\br}{{\bf r}}
\newcommand {\gvt}{{\gamma v_2}}

\newcommand {\xpl}{{\hat{\bf x}_{_\parallel}}}

\newcommand {\vq}{\overline{|{\bf v}_{_\perp}({\bf q})|^2}}

\newcommand {\rsv}{\overline{|{\bf v}_{_\perp}({\bf r})|^2}}

\newcommand {\red}{\texttt{\color{red}}}
\newcommand {\blue}{\texttt{\color{blue}}}
\newcommand{\vp}{{\bf v}_{_\perp}}
\newcommand{\vt}{{\bf v}_{_T}}
\newcommand{\qp}{{\bf q}_{_\perp}}
\newcommand{\bv}{{\bf v}}
\newcommand{\brp}{{\bf r}_{_\perp}}
\newcommand{\rp}{r_{_\perp}}
\newcommand{\chip}{\chi_{_\perp}}

\newcommand{\chipar}{ \xi_{_\parallel}}
\newcommand{\rpar}{ r_{_\parallel}}
\newcommand{\qpar}{ q_{_\parallel}}
\newcommand{\mqp}{ q_{_\perp}}

\newcommand {\fl}{\cite{TT1, TT2,TT3,TT4, NL}}

\def\lsim{\:\raisebox{-0.5ex}{$\stackrel{\textstyle<}{\sim}$}\:}

\def\gsim{\:\raisebox{-0.5ex}{$\stackrel{\textstyle>}{\sim}$}\:}

\def\Dtens{\mbox{\sffamily\bfseries D}}

\def\Wtens{\mbox{\sffamily\bfseries W}}

\def\Ptens{\mbox{\sffamily\bfseries P}}

\def\Otens{\mbox{\sffamily\bfseries O}}

\def\TT{\cite{TT1,TT2,TT3,TT4, NL}}

\def\Qtens{\mbox{\sffamily\bfseries Q}}

\def\Q{\mbox{\sffamily\bfseries Q}}

\def\Ntens{\mbox{\sffamily\bfseries N}}

\def\Ctens{\mbox{\sffamily\bfseries C}}

\def\Itens{\mbox{\sffamily\bfseries I}}

\def\Atens{\mbox{\sffamily\bfseries A}}

\def\A{\mbox{\sffamily\bfseries A}}

\def\Ktens{\mbox{\sffamily\bfseries K}}

\def\Vtens{\mbox{\sffamily\bfseries V}}

\def\Gtens{\mbox{\sffamily\bfseries G}}

\def\ftens{\mbox{\sffamily\bfseries f}}

\def\vtens{\mbox{\sffamily\bfseries v}}

\def\nabbold{\mbox{\boldmath $\nabla$\unboldmath}}

\def\nabvec{\mbox{\boldmath $\nabla$}}

\def\sigtens{\mbox{\boldmath $\sigma$\unboldmath}}

\def\etatens{\mbox{\boldmath $\eta$\unboldmath}}

\def\beq{\begin{equation}}

\def\bea{\begin{eqnarray}}

\def\eeq{\end{equation}}

\def\eea{\end{eqnarray}}

\title{Swarming in the Dirt: Ordered Flocks with Quenched Disorder}
\author{John Toner}
\author{Nicholas Guttenberg}

\affiliation{Institute for Theoretical Science and  Department of Physics, University of Oregon, Eugene, OR 97403}
\author{Yuhai Tu}
\affiliation{IBM T. J. Watson Research Center, Yorktown Heights, NY 10598}

\begin{abstract}

The effect  of  quenched (frozen) 
disorder on the collective motion of active particles is analyzed. We find that 
active polar systems are far more robust against quenched disorder than 
equilibrium ferromagnets.  
Long ranged order (
a non-zero average velocity $\langle{\bf v}\rangle$) persists in the presence of quenched disorder even in spatial dimensions $d=3$;  
in $d=2$, quasi-long-ranged order (i.e., spatial velocity correlations that decay as a power law with distance) occurs.  In equilibrium systems, only quasi-long-ranged order in $d=3$ and short ranged order in $d=2$ are possible. 
Our theoretical predictions for two dimensions are borne out by  simulations.  

 \end{abstract}


\maketitle

{\bf Introduction.} A great deal of the immense current interest  in ``Active Matter" focuses on  coherent collective motion, i.e.,  ``flocking" \cite{boids, Vicsek, TT1,TT2,TT3,TT4, NL}, or  
``swarming" \cite{dictyo,rappel1}. 
Such coherent motion occurs over 
a wide range 
 of length scales:  from macroscopic organisms
 to mobile  macromolecules in living cells \cite{dictyo,rappel1,actin,microtub}.
Such coherent motion is possible even in $d=2$ \cite{Vicsek},  in  apparent violation of  the Mermin-Wagner theorem
\cite{MW}. This has been explained by the ``hydrodynamic" theory of flocking \cite{TT1,TT2,TT3,TT4, NL}, which shows that, unlike equilibrium ``pointers", non-equilibrium ``movers" {\it can}
spontaneously break a continuous symmetry (rotation invariance) by developing long-ranged orientational order  (as they must  to have a  non-zero average velocity $\left<{\bf v} ({\bf r}, t) \right>\ne \bf 0$), even in noisy systems with  only short ranged interactions in spatial dimension $d=2$, and 
in flocks with birth and death \cite{Malthus}. In equilibrium systems, 
even {\it arbitrarily weak} quenched random fields  destroy long-ranged ferromagnetic order in all spatial dimensions $d\le4$ \cite{Harris, Geoff, Aharonyrandom,Dfisher}. This
raises the question:
can the non-linear, non-equilibrium effects that make long-ranged order possible in 2d flocks   without quenched disorder stabilize them   when random field  disorder is present? 
Simulations of flocks with quenched disorder 
\cite{Peruani,Das}
find  quasi-long-ranged order in $d=2$; that is:  
\begin{equation}
\overline{{\bf v}({\bf r},t)\cdot{\bf v}({\bf r}^{~\prime},t)}\propto|{\bf r}-{\bf r}^{~\prime}|^{-\eta}~,
\label{qlro}
\end{equation}
where the exponent $\eta$ is non-universal (that is, system dependent), and the overbar denotes an average over $\br$ with fixed ${\bf r}-{\bf r}^{~\prime}$.

In this paper and the accompaning long paper (ALP), we address this problem analytically and by   simulations. The analytical approach (the focus of this paper) extends the hydrodynamic theory of flocking developed in \cite{TT1,TT2,TT3,TT4, NL} to include quenched disorder. 
Both approaches confirm  that   flocks are more robust against quenched disorder than  ferromagnets.  
Specifically, we find that flocks {\it can} develop long ranged order in three dimensions, and quasi-long-ranged order 
in two dimensions, due to strong non-linear effects, in contrast to the equilibrium case, in which only short-ranged order is possible in  
two  dimensions \cite{Harris, Geoff, Aharonyrandom,Dfisher}, and only quasi-long-ranged order in three dimensions. 
We also determine exact scaling laws for velocity fluctuations for one range of hydrodynamic parameters in $d=3$.

{\bf The hydrodynamic theory.}  To study the effects of quenched disorder for flocking, we use the hydrodynamic theory of \cite{TT1,TT2,TT3,TT4,NL}, modified only by the inclusion of a quenched random force {\bf f}.
 In the ordered phase, this takes the form \cite{NL} of the following pair of coupled equations of motion for the fluctuation ${\bf v}_{\perp}(\vec{r},t)$ of the local velocity of the flock perpendicular to the direction of mean flock motion (which mean direction will hereafter denoted as "$\parallel$"), and the departure $\delta\rho(\vec{r},t)$ of the density from its mean value $\rho_0$:


\begin{widetext}
\begin{eqnarray}
&\partial_{t} {\bf v}_{\perp} + \gamma\partial_{\parallel} 
{\bf v}_{\perp} + \lambda \left({\bf v}_{\perp} \cdot
{\bf \nabla}_{\perp}\right) {\bf v}_{\perp} =-g_1\delta\rho\partial_{\parallel} 
{\bf v}_{\perp}-g_2{\bf v}_{\perp}\partial_{\parallel}
\delta\rho-g_3{\bf v}_{\perp}\partial_t
\delta\rho -{c_0^2\over\rho_0}{\bf \nabla}_{\perp}
\delta\rho -g_4{\bf \nabla}_{\perp}(\delta \rho^2)\nonumber\\&+
D_B{\bf \nabla}_\perp\left({\bf \nabla}_\perp\cdot{\bf v}_\perp\right)+
D_T\nabla^{2}_{\perp}{\bf v}_{\perp} +
D_{\parallel}\partial^{2}_{\parallel}{\bf v}_{\perp}+\nu_t\partial_t{\bf \nabla}_{\perp}\delta\rho+\nu_\parallel\partial_\parallel{\bf \nabla}_{\perp}\delta\rho+{\bf f}_{\perp} ,
\label{vEOMbroken}\\
&\partial_t\delta
\rho +\rho_o{\bf \nabla}_\perp\cdot{\bf v}_\perp
+\lambda_{\rho}{\bf \nabla}_\perp\cdot({\bf v}_\perp\delta\rho)+v_2
\partial_{\parallel}\delta
\rho =D_{\rho\parallel}\partial^2_\parallel\delta\rho+D_{\rho v} \partial_{\parallel}
\left({\bf \nabla}_\perp \cdot {\bf v}_{\perp}\right)+\phi\partial_t\partial_\parallel\delta\rho
+\partial_\parallel(w_1 \delta\rho^2+w_2|{\bf v}_\perp|^2),  
\label{cons broken}
\end{eqnarray}
\end{widetext}
where  $\gamma$, $\lambda$,  $\lambda_\rho$,
$c_0^2$, $g_{1,2,3,4}$, $w_{1,2}$,  $D_{B\rm{eff},T,\parallel, \rho\parallel, \rho v}$, $\nu_{t,\parallel}$, $v_2$, $\phi$,  and $\rho_0$  are all phenomenological constants.


To treat quenched disorder, we simply take this random force to be {\it static}; i.e., to  depend {\it only} on position: ${\bf f}({\bf r},t)={\bf f}({\bf r})$, and not on time $t$ at all, with short-ranged spatial  correlations:

\begin{equation}
 \overline{ f^\perp_i ({\bf r}) f^\perp_j ({\bf r}^{~\prime})} = \Delta \delta^\perp_{ij} \delta^d ({\bf r}-{\bf r}^{~\prime})
\label{fcor}
\end{equation}
where the overbar  denotes averages over the quenched disorder, and $\delta^\perp_{ij}=1$ if and only if $i=j\ne\parallel$, and is zero for all other $i$, $j$.  We will also assume ${\bf f_\perp}$ is zero mean, and Gaussian. 

{\bf The linearized hydrodynamic theory and anisotropic fluctuations.}  
Our first step in analyzing these equations is to linearize them. We then Fourier transform them in space and time, and decompose 
the velocity ${\bf v}_{_\perp}$  
along and perpendicular to the projection ${\bf q}_{_\perp}$ of $\bq$  perpendicular to the mean direction of flock motion: 
$v_L\equiv {\bf v}_{_\perp}\cdot{\bf q}_{_\perp}/q_{_\perp}$, 
$
{\bf v}_{_T}\equiv {\bf v}_{_\perp}- v_L {{\bf q}_{_\perp}\over q_{_\perp}}
$.  Note that the 
``transverse" velocity  $\vt$ does not exist in $d=2$, where there are no directions that are orthogonal to both $\qp$ and the mean direction of flock motion $\xpl$. This has important consequences,  
as we will see later. 

The   set of
coupled linear algebraic equations  
for $\delta\rho$, ${\bf v}_{T}$, and $v_L$ that we thereby obtain can be solved analytically to obtain the strength of the fluctuations (details are given in the ALP):

\begin{equation}
\overline{|v_L({\bf q})|^2}= \frac{(\tilde{\Delta}\cos^2\theta_{{\bf q}})q^{-2}}{\epsilon^2(\theta_{{\bf q}})q^2 + (\sin^2\theta_{{\bf q}}-\left[{\gamma v_2\over c_0^2}\right] \cos^2\theta_{{\bf q}})^2},
\label{vLanglefluc}
\end{equation}
\begin{equation}
\overline{|\delta\rho({\bf q})|^2}=\frac{[\tilde{\Delta}(\rho_0^2/v_2^2)\sin^2\theta_{{\bf q}}]q^{-2}}{\epsilon^2(\theta_{{\bf q}})q^2 + (\sin^2\theta_{{\bf q}}-\left[{\gamma v_2\over c_0^2}\right]\cos^2\theta_{{\bf q}})^2}~,
\label{rhoanglefluc}
\end{equation}
and
\begin{equation}
\overline{|{\bf v}_T({\bf q})|^2}={(d-2)\Delta\over \gamma^2q^2\left[\epsilon_T^2(\theta_{{\bf q}})q^2 +  \cos^2\theta_{{\bf q}}\right]}~,
\label{vTanglefluc}
\end{equation}
with $q$ the magnitude of the wavevector ${\bf q}$ and $\theta_{{\bf q}}$ the angle   between ${\bf q}$ and the direction $\hat{x}_\parallel$ of mean flock motion.  In Eqs. (\ref{vLanglefluc}-\ref{vTanglefluc}), $\tilde\Delta\equiv{v_2^2\Delta\over c_0^4}$ and $\epsilon(\theta_{{\bf q}})$ and $\epsilon_T(\theta_{{\bf q}})$ are the finite direction-dependent damping coefficients (see ALP for their expressions).

From Eqs. (\ref{vLanglefluc}-\ref{vTanglefluc}), we immediately see that there is an important distinction between the cases  $\gvt>0$ and $\gvt<0$. In the former case,  fluctuations of $v_L$ and $\rho$ are highly anisotropic: 
they scale like $q^{-2}$ for all directions of ${\bf q}$ {\it except}  when $\theta_{{\bf q}}=\theta_c$ or $\pi-\theta_c$,  where we have defined a critical angle of propagation $\theta_c\equiv\arctan \left[{\sqrt{\gamma v_2}\over c_0}\right]$.   For these special directions (which only exist if $\gvt>0$) both $\overline{|v_L({\bf q})|^2}$ and $\overline{|\delta
\rho({\bf q})|^2}$ scale like $q^{-4}$. On the other hand, when $\gvt<0$, fluctuations of $v_L$ and $\rho$ are essentially isotropic: 
they scale as $q^{-2}$ for {\it all} directions of ${\bf q}$.


Fluctuations of $\vt$, however, are {\it always} anisotropic, diverging as $q^{-4}$ for $\theta_{{\bf q}}=\pi/2$, and as $q^{-2}$ for all other directions of $\bq$. Of course, there {\it are} no such fluctuations in $d=2$, since, as noted earlier, $\vt$ does not exist in that case, as reflected by the factor of $(d-2)$ in Eq. (\ref{vTanglefluc}).

These special directions ($\theta_c$ and ${\pi\over 2}$) dominate the real space fluctuations $\overline{|{\bf v}_\perp({\bf r})|^2}$  and $\overline{|\delta\rho({\bf r})|^2}$, which can be obtained by integrating
$\overline{|\delta\rho({\bf q})|^2}$, $\overline{|v_L({\bf q})|^2}$, and $\overline{|{\bf v}_T({\bf q})|^2}$ over all wavevector ${\bf q}$. In particular, we have: 
\begin{equation}
\overline{|{\bf v}_\perp({\bf r})|^2}=\int q^{d-1}dq\int d\Omega_\bq\left(\overline{|{\bf v}_T({\bf q})|^2}+\overline{|v_L({\bf q})|^2}\right),
\label{vrflucscale}
\end{equation}
where $\int d\Omega_\bq$ denotes an integral over the directions of $\bq$. As shown in the ALP, this angular integral scales like $q^{-3}$ for the $\vt$ term in Eq.~(\ref{vrflucscale}), except, of course, in $d=2$, where that term does not exist. The $v_L$ term also scales like $q^{-3}$ when $\gvt>0$, due to the aforementioned divergence of $\overline{|v_L({\bf q})|^2}$ as $\theta_{_{\bq}}\to\theta_c$. However, it only scales like $q^{-2}$ when $\gvt<0$, since $\overline{|v_L({\bf q})|^2}$ does not blow up for any direction of $\bq$ in that case.

Hence, if {\it either} $d>2$, {\it or} $\gvt>0$, Eq.~(\ref{vrflucscale}) implies
\begin{equation}
\overline{|{\bf v}_\perp({\bf r})|^2}\propto\int q^{d-4}dq~ 
\ ,
\label{vrflucscale2}
\end{equation}
which 
clearly diverges in the long wavelength (i.e., infra-red, or $q\rightarrow 0$) limit for $d\le 3$. Thus, according to the linearized theory, there should be no long-ranged orientational order (a nonzero $\overline{{\bf v}({\bf r})}$)  for $d\le 3$, no matter how weak the disorder. In the critical dimension $d=3$, quasi-long-ranged order (with algebraic decay of velocity correlations in space), should, again according to the {\it linearized} theory, occur.

However, for the case $\gvt<0$ (when $\overline{|v_L({\bf q})|^2}$ has no soft directions) and $d=2$ (when $\vt$ does not exist), we have
\begin{equation}
\overline{|{\bf v}_\perp({\bf r})|^2}\propto\int q^{d-3}dq~ 
\ ,
\label{vrflucscalespecial}
\end{equation}
which only diverges in $d\le2$. In $d=2$, this divergence is only logarithmic, suggesting quasi-long-ranged order characterized by Eq.~(\ref{qlro}).



We thus see that
there is a significant difference between dimension $d=2$ and $d>2$, and between $\gamma v_2>0$ and $\gamma v_2<0$. 
Thus there are  four distinct cases of physical interest.  The linear theory just presented predicts
quasi-long-ranged order for three of these four cases: $d=3$ for both $\gamma v_2>0$ and $\gamma v_2<0$, and  $d=2$ for the case  $\gamma v_2<0$. For the remaining case, $d=2$ and $\gamma v_2>0$, the linear theory predicts only short-ranged order.

However, in  the full, non-linear theory, there is true long-ranged order - specifically, a  non-zero average velocity $\overline{{\bf v} ({\bf r}, t)}\ne \bf 0$  for $d=3$, and quasi-long-ranged order in $d=2$, in both cases $\gamma v_2>0$ and $\gamma v_2<0$. Below, we will present the detailed analysis of the full non-linear model 
for  the simplest case: $\gamma v_2<0$ and $d>2$, and for the case we simulate, $\gamma v_2>0$ and $d=2$. We defer detailed discussion of the other two cases to the ALP.


{\bf Breakdown of the linearized theory in $d\le 5$.} 
We now show that the non-linearities explicitly displayed in the coarse-grained equations of motion 
radically change the scaling of fluctuations in flocks with quenched disorder for all spatial dimensions $d\le5$. Furthermore, this change in scaling stabilizes orientational order, i.e., makes it possible for the flock to acquire a non-zero mean velocity
($<{\bf v}>\ne 0$)) in three dimensions.


We begin by demonstrating  this for $d\ne2$ and $\gvt<0$ by power counting .  (The same conclusion also holds for   $\gvt>0$, but we defer the more complicated argument for that case to the ALP.) 
Due to the anisotropy, we rescale coordinates $r_\parallel$ along the  direction of flock motion differently from
those ${\bf r_\perp}$ orthogonal to that direction, and also rescale time and the fields: 
\begin{eqnarray}
&&{\bf r}_\perp \rightarrow b {\bf r}_\perp \ , ~~
r_\parallel\rightarrow b^\zeta r_\parallel  \ , ~~
t \rightarrow b^z t  \ , ~~
{\bf v}_\perp \rightarrow b^\chi {\bf v}_\perp   \ , ~~  \nonumber\\
&&\delta \rho \rightarrow b^{\chi_\rho} \delta \rho \ .
\label{rescale}
\end{eqnarray}

These rescalings  
relate  the parameters in the rescaled equations (denoted by primes) are related to those of the unrescaled equations. We will focus on the parameters $\Delta$, $\gamma$, and $D_T$, and the combination of parameters ${c_0^2\over\rho_0}$, which control the fluctuations in the dominant direction $\theta_\bq=\pi/2$  of wavevector  ${\bf q}$. We easily find:
\begin{eqnarray}
&&\gamma^{\prime}=b^{z-\zeta}\gamma \ , ~~\Delta^{\prime}=b^{2(z-\chi)+1-d-\zeta}\Delta \ ,
\label{gammaDeltarescale}
\\
&&\left({c_0^2\over\rho_0}\right)^{\prime}=b^{{\chi_\rho-\chi+z-1}}\left({c_0^2\over\rho_0}\right)  \ , ~~
D_T^{\prime}=b^{z-2}D_T \ . ~~
\label{Drescale}
\end{eqnarray}
We can thus keep the scale of the fluctuations 
fixed by choosing the exponents $z$, $\zeta$, $\chi$, and $\chi_\rho$ to obey
\begin{eqnarray}
&&z-\zeta=0  \ , ~~
\chi_\rho-\chi+z-1=0 \ , ~~
z-2=0 \ ,~~\nonumber\\
&&2(z-\chi)+1-d-\zeta=0 \ .
\label{linfix}
\end{eqnarray}
Solving these yields
\begin{eqnarray}
z_{\rm{lin}}=\zeta_{\rm{lin}}=2  ~,~\chi_{\rm{lin}}={3-d\over2} ~,~\chi_{\rho,\rm{lin}}={1-d\over2} ~~.
\label{linexp}
\end{eqnarray}
The subscript ``lin" in these expressions denotes the fact that we have determined these exponents ignoring the effects of the non-linearities in the equations of motion (\ref{vEOMbroken}) and (\ref{cons broken}). We now use them to determine in what spatial dimension $d$ those non-linearities become important.

Upon the rescalings (\ref{rescale}), the  non-linear terms
$\lambda$,  and
$g_{1,2,3,4}$ in the $\vp$ equation of motion (\ref{vEOMbroken})  obey
\begin{eqnarray}
\lambda^{\prime}=b^{z+\chi-1}\lambda= b^{5-d\over2}\lambda \ ,
\label{lambdarescale}
\\
g_{1,2,3,4}^{\prime}=b^{z+\chi_\rho-\zeta}g_{1,2,3,4} =b^{1-d\over2}g_{1,2,3,4} \ .
\label{nlrescale}
\end{eqnarray}


By inspection of Eq.~(\ref{nlrescale}), we see that only $\lambda$ becomes relevant in any spatial dimension $d>1$; in fact, it becomes relevant for $d\le d_c=5$. The $g_i$'s are all irrelevant, and can be dropped. Furthermore, if we restrict ourselves to consideration of the transverse modes $\vt$, which we can do by projecting 
the spatial Fourier transform of Eq.~(\ref{vEOMbroken}) perpendicular to $\qp$, we see that there is {\it no} coupling between $\vt$ and $\rho$ {\it at all}, even at nonlinear order. Hence, $\rho$ completely drops out of the problem of determining the fluctuations of $\vt$. And since $\vt$ is, as we saw in our treatment of the linearized version of this problem, the dominant contribution to the velocity fluctuations when $d>2$ (so that $\vt$ actually exists) and $\gamma v_2<0$ (so that there is no direction of $\bq$ for which the {\it longitudinal} velocity fluctuations $v_L$ diverge more strongly than $1/q^2$ in the linearized approximation), this means that the long distance scaling of the velocity fluctuations will be the same as in a model with no density fluctuations {\it at all}; that is, an incompressible model, in which 
$\nabla_\perp\cdot\vp=0.$

We now note two useful facts: 

\noindent 1) The only nonlinearity (the $\lambda$ term) can  be written as a total $\perp$-derivative. This follows from the 
identity:
\begin{eqnarray}
\left({\bf v}_{_\perp} \cdot
{\bf \nabla}_{_\perp}\right) v^{\perp}_i=\partial^\perp_j\left(v^\perp_jv^\perp_i\right)-v^\perp_i{\bf \nabla}_{_\perp}\cdot{\bf v_{_\perp}}\,\, .
\label{trans1}
\end{eqnarray}
The first term on the right hand side of this expression is obviously a total $\perp$-derivative. The second term vanishes since $\nabla_\perp\cdot\vp=0$, which implies that the nonlinearity can {\it only}
renormalize terms which 
involve $\perp$-derivatives (i.e., $D_T^0$); specifically, there are {\it no} graphical corrections to either $\gamma$ or $\Delta$.

\noindent 2)  There are no graphical corrections for
$\lambda$ either, because the equations of motion  (\ref{vEOMbroken}) and (\ref{cons broken}) have  an exact
``pseudo-Galilean invariance" symmetry \cite{pseudo}, i.e., they remain
unchanged by
a pseudo-Galilean transformation:
\begin{eqnarray}
{\bf r}_\perp \to {\bf r}_\perp-\lambda {\bf v}_1 t~~~,~~
{\bf v}_{_\perp} \to {\bf v}_{_\perp} + {\bf v}_1~~~
,\label{Gal}
\end{eqnarray}
 for arbitrary constant vector ${\bf v}_1\perp\hat{x}_{_\parallel}$.
Since such an exact symmetry  must continue to hold upon renormalization, with the {\it same} value of  $\lambda$,  the parameter $\lambda$ cannot be graphically renormalized.

Taken together, these two facts imply that Eq.~(\ref{gammaDeltarescale}) and the first equality of Eq.~(\ref{lambdarescale}) are exact, even when graphical correction are included.
Therefore, to get a fixed point, we must have
\beq
z-\zeta=0 ~ , ~ 2(z-\chi)+1-d-\zeta=0 ~ , ~ z+\chi-1=0 ~ ,
\label{expcond}
\eeq
which imply
\begin{eqnarray}
z={d+1\over 3}=\zeta  \ ,~~
\chi={2-d\over 3}  \ .
\label{zchiexact1}
\end{eqnarray}
 
The fact that $\chi<0$ for all $d$ in the range $2<d<5$ implies that velocity fluctuations get smaller as we go to longer and longer length scales; this implies the existence of long ranged order (i.e., a non-zero average velocity $\overline\bv\ne{\bf 0}$) in all of those spatial dimensions. The physically realistic case in this range is, of course, $d=3$.

These exponents imply that Fourier transformed velocity correlations take the form:
\begin{eqnarray}
\overline{|{\bf v}_{_\perp}({\bf q})|^2}={h\left({\qpar/\Lambda
 \over
(\mqp/\Lambda)^\zeta }
\right)\over \qpar^{2} }
\propto\left\{
\begin{array}{ll}
\mqp^{-2\zeta},
&\left({\mqp\over\Lambda}\right)^\zeta\gg{\qpar\over\Lambda}\,,
\\ \\
\qpar^{-2},
&\left({\mqp\over\Lambda}\right)^\zeta\ll{\qpar\over\Lambda}\ ,
\\ \\
\end{array}\right.
\label{vsc2}
\end{eqnarray}
where $\Lambda$ is an ultraviolet cutoff.

{\bf Nonlinear effects for $\gamma v_2>0$, $d=2$. }  Now ``longitudinal" fluctuations (i.e., $\delta\rho$ and $v_{_L}$) become important, which causes the $g$ and $w$ non-linearities in the equations of motion (\ref{vEOMbroken}) and (\ref{cons broken}) important. This  prevents us from  making such a compelling argument for exact exponents. However, our experience with the annealed noise problem suggests a way forward. In that annealed case, the {\it assumption} that below the critical dimension only {\it one} of the non-linearities, namely the convective $\lambda$ term, is  relevant, makes it  possible to determine exact exponents in $d=2$. These exponents agree extremely well with 
simulations of flocking \cite{TT1,TT2,TT3,TT4,NL}. Thus  this assumption appears to be correct for the annealed problem, which suggests that it might also be true in the quenched disorder problem.

If it is, then the two points that we used to determine the exact exponents for the $\gvt<0$, $d\ne2$ case just considered also hold here. In this case, the $\lambda$ non-linearity can be written as a total derivative 
because $\vp$ has only one component in $d=2$, so $\left({\bf v}_{_\perp} \cdot
{\bf \nabla}_{_\perp}\right) v^{\perp}_i=v_\perp\partial_{_\perp}v_{_\perp}=\partial_{_\perp}(v_{_\perp}^2/2)\,\,.$ Pseudo-Galilean invariance also applies once $\lambda$ is the only relevant non-linearity \cite{pseudo}. 

Hence, the arguments we made earlier for the exact exponents for the case $\gamma v_2<0$, $d\ne2$ also apply for $\gamma v_2>0$, $d=2$. This implies that the exponents of Eq.~(\ref{zchiexact1}) apply here as well, albeit with $d=2$, which implies $z=\zeta=1$, $\chi=0$. The vanishing of $\chi$ implies quasi-long-ranged order (Eq. (\ref{qlro})), while the fact that $\zeta=1$ implies that fluctuations scale isotropically. Note that this is in strong contradiction to the linear theory, which predicts extremely {\it anisotropic} scaling of fluctuations when $\gvt>0$, as it is here.

\begin{figure}
\includegraphics[trim={7.0cm 3cm 7.0cm 4cm}, clip=true,width=2.75in]{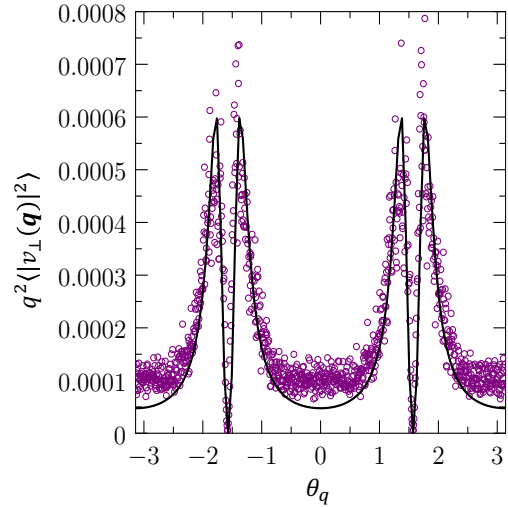}
\caption{Fourier-space velocity-velocity correlation function of a Vicsek flock as a function of the direction of wavevector $\bq$ in the presence of quenched disorder. The solid line is the theoretical prediction from the continuum hydrodynamic theory. The linearized theory predicts that $q^2\overline{ |v_{\perp}({\bf q})|^2 }$  diverges as $\bq\rightarrow\bf{0}$ at some non-universal  critical angle $\theta_c$, while the non-linear theory predicts that the divergence at $\theta_c$ will be cut off, leaving a large, but finite, maximum. In our simulations, $\theta_c \approx 78^\circ$.}
\label{Fig3Correlation}
\end{figure}

The physical origin of this restoration of isotropic scaling is that the damping coefficient  $\epsilon^2(\theta_{{\bf q}})$ is renormalized by non-linear fluctuation effects by an amount that scales like $q^{-2}$ as $\bq\to{\bf 0}$ and $\theta_\bq\to\theta_c$, cancelling off the explicit $q^2$ in Eq.~(\ref{vLanglefluc}), and thereby making the fluctuations scale isotropically. Since this nonlinear effect is caused by disorder induced fluctuations, we expect the finite $\bq\to{\bf 0}$ limiting value of $q^2\epsilon^2(\theta_{{\bf q}})$, which we define as $\delta$ (i.e., $\delta \equiv {\lim\over q\to 0}  q^2\epsilon^2(\theta_c)$) to get very small as the strength $\Delta$ of the disorder does. Since our simulations are done at weak disorder, we expect $\delta$ to be small, which implies a sharp peak in a plot of $q^2\overline{|{\bf v}_{_\perp}({\bf q})|^2}$ versus $\theta_\bq$. Specifically, our analysis implies
\begin{equation}
q^2 \overline{ |v_{\perp}({\bf q})|^2 } \propto \frac{\Delta  \cos^2(\theta)}{[\sin^2(\theta)-\tan^2(\theta_c) \cos^2(\theta)]^2+\delta} \ ,
 \label{vq2}
\end{equation}
which is confirmed by our simulations as illustrated in 
Fig.~\ref{Fig3Correlation}  (described in detail in the ALP).

It should be noted here, however, that this result cannot continue to hold down to arbitrarily small $\bq$. This is because quasi-long-ranged order, which Eq.~(\ref{vq2}) (or, equivalently, our result $\chi=0$) implies, is inconsistent with macroscopic anisotropy. Therefore, at large enough length scales that the velocity correlation function in Eq.~(\ref{qlro}) becomes $\ll (\overline{\bv(\br)})^2$, isotropy must be restored, so our strongly anisotropic result Eq.~(\ref{vq2}) must break down.
We believe that what happens here is the same as in an equilibrium two-dimensional nematic \cite{2dnem}, in which 
isotropy is restored by  slow (logarithmic) effects 
that our scaling argument above is too crude to pick up.
 
We also note that those subtle effects should only become apparent on length scales that grow like $\exp[\rm{constant}/\Delta]$ for small $\Delta$, which will become astronomically large length scales if the noise strength $\Delta$ is small, as it  is in our simulations. This appears to be the case in our simulations, since, as shown in Fig. ~\ref{Fig3Correlation}, they still exhibit considerable anisotropy.

{\bf Summary.} We have studied a fully nonlinear hydrodynamic equation for flocking in the presence of quenched disorder. We find that the critical dimension for the nonlinear terms to become relevant is $d_c=5$. For $d<5$ and the combination of phenomenological parameters $\gvt<0$ 
we determine of all the scaling exponents Eq.~(\ref{zchiexact1}).   These predicted exponents show that flocks with non-zero quenched disorder can still develop long ranged order in three dimensions, and quasi-long-ranged order in two dimensions, in strong contrast to the equilibrium case, in which any amount of quenched disorder destroys ordering in both in two  and three dimensions \cite{Harris, Geoff, Aharonyrandom}. This prediction is consistent  with the simulation results of Chepizhko et. al. \cite{Peruani} and Das et al. \cite{Das} and ourselves. 
(see ALP for more comparisons). 

{\bf Acknowledgements.} We are very grateful to F. Peruani for invaluable discussions of his work on this subject, and to R. Das, M. Kumar, and S. Mishra for sharing their results with us prior to publication. JT thanks Mike Cates for pointing out the possibility that $\gamma v_2$ might be negative, and that this could affect our results. He also thanks the IBM T. J. Watson Research Center, Yorktown Heights, New York; the Institut f\"{u}r Theoretische Physik II: Weiche Materie,
Heinrich-Heine-Universit\"{a}t, D\"{u}sseldorf; the Max Planck Institute for the Physics of Complex Systems  Dresden; the Department of Bioengineering at Imperial College, London; The Higgs Centre for Theoretical Physics at the University of Edinburgh; the Kavli Institute for Theoretical Physics,University of California, Santa Barbara; and the Lorentz Center of Leiden University; for their hospitality while this work was underway.



\begin{thebibliography}{000}
\bibitem{pseudo} In order for {\it both} equations of motion ((\ref{vEOMbroken}) and (\ref{cons broken})) to
have this pseudo-Galilean invariance, it is actually also necessary that $w_1=\lambda$.
However, for $\gvt<0$, $w_1$ is irrelevant, so we can choose $w_1=\lambda$ without changing the large distance scaling. For the $\gvt>0$, $d=2$ case, we will show in the ALP
that the existence of a pseudo-Galilean invariance for one value of $w_1$ ensures the same relation between exponents for all values of $w_1$ as that obtained when $w_1=\lambda$.
\end{thebibliography}

\end{document}